# MAXIMIZING STRENGTH OF DIGITAL WATERMARKS USING FUZZY LOGIC

Sameh Oueslati[1] and Adnane Cherif[1] and Bassel Solaiman[2]

[1]Department of Physics: Laboratory of Signal Processing, University of Sciences of Tunis 1060, Tunisia
`sameh.oueslati@telecom-bretagne.eu`
`adnane.cher@fst.rnu.tn`

[2]Department of Image and Information Processing, Higher National School of Telecommunication of Bretagne Technopôle, Brest-Iroise, CS 83818, 29238 Brest Cedex 3 France.
`Basel.Solaiman@telecom-bretagne.eu`

## ABSTRACT

*In this paper, we propose a novel digital watermarking scheme in DCT domain based fuzzy inference system and the human visual system to adapt the embedding strength of different blocks. Firstly, the original image is divided into some 8×8 blocks, and then fuzzy inference system according to different textural features and luminance of each block decide adaptively different embedding strengths. The watermark detection adopts correlation technology. Experimental results show that the proposed scheme has good imperceptibility and high robustness to common image processing operators.*

## 1. INTRODUCTION

The necessity of fast and secure diagnosis is vital in the medical world. Nowadays, the transmission of visual data is a daily routine and it is necessary to find an efficient way to transmit them over networks [13], [15], [17]. The main objective is to guarantee the protection of medical images during transmission, and also once this digital data is archived [6], [7]. The subsequent challenge is to ensure that such coding withstands severe treatment such as compression [26]. When a physician receives a visit from a patient, he often requires a specialist opinion before giving a diagnosis. One possible solution is to send images of the patient, along with a specialist report, over a computer network [33], [1]. Nevertheless, computer networks are complex and espionage is a potential risk. We are therefore faced with a real security problem when sending data. Forethical reasons, medical imagery cannot be sent when such a risk is present, and has to be better protected [11]. Watermarking is the best form of protection in cases such as this. Many different techniques for the watermarking of text already exist. Recently, watermarking has been proposed for medical information protection. Even though most of the work on watermarking has concerned medical images in order to verify image integrity or improve confidentiality [32], watermarking also provides a new way to share data. Basically, watermarking is defined as the invisible embedding or insertion of a message in a host document, an image, for example. Anand *et al*. [9], proposed to insert an encrypted version of the electronic patient record (EPR) in the LSB (Least Significant Bit) of the gray scale levels of a medical image. Although the degradation in the image quality is minimum, the limitations and fragility of LSB watermarking schemes is well-known. Miaou *et al*. [23] proposed a method to authenticate the origin of the transmission, the message embedded is an ECG, the diagnosis report and physician's information. Macq and Dewey [24] insert information in the headers of medical images. These approaches are not robust against attacks such as filtering, compression, additive noise, etc. neither to geometrical attacks such as rotation or scaling transformations. In this work, we propose the use of DICOM metadata as a watermark to embed in medical images extracted from the DICOM file. We introduce a metric of an image quality evaluation ($wPSNR$). This distortion metric, that has no relation with the content characteristics of the image fits to the HVS and therefore is more suitable for digital Watermarking. In this method The FIS and the HVS combined are used to adjust the watermarking strength. In order for the embedded watermark to be even more robust against different types of attacks, it is essential to add as powerful invisible watermark as possible. Based on the above reasons, the proposed scheme in this paper divides the original image into some 8×8 blocks, Feature Extraction by Co-occurrence Matrix, and the FIS according to different textural features and luminance of each block decide adaptively different embedding strengths. Finally, the watermark extraction is semi-blind, i.e., it is accomplished without using the original image which makes the approach higher security. As a result, the watermark is more robust and imperceptible. The remainder of the paper is organized as follows. Section 2 provides a detailed a description of the HVS model and FIS, Section 3 describes the method for embedding and extraction watermark. In Section 4, the experimental results and comparisons are shown. The conclusions of our study are stated in Section 5.

## 2. HUMAN VISUAL SYSTEM AND TEXTURAL FEATURES SELECTION
### 2.1. Human Visual System (HVS) in Watermarking

It is known that there is a trade off between the imperceptibility and robustness of a digital watermarking system. Up to now, the watermark's invisibility issue is only tackled by the embedding depth [22], [10]. But, this action is very limited because the watermark signal, once mapped in the media space, is spread all over the image. Uniform areas of the image are very sensitive to watermark addition so they only support extremely small embedding depth, whereas edge areas, for instance, support deeper watermark addition. The new spatial masking is built according to the image features such as the brightness, edges, and region activities. With the same watermark embedding, the quality of watermarked image using the proposed adaptive masking is much better than the one without using the adaptive masking. The human visual system has been characterized with several phenomena that permit pixel element adjustments to elude perception.

### 2.2. Feature Extraction by Co-occurrence Matrix

A general procedure for extracting textural properties of image was presented by Haralick *et al*. [2]. Each textural feature was computed from a set of $COM$ probability distribution matrices for a given image. The COM measures the probability that a pixel of a particular grey level occurs at a specified direction and a distance from its neighbouring pixels. Formally, the elements of a $G \times G$ gray level co-occurrence matrix $P_d$ for a displacement vector $d = (dx, dy)$ is defined as:

$$P_d(i, j) = |\{((r,s),(t,v)) : I(r,s) = i, I(t,v) = j\}| \quad (1)$$

Where $I(.,.)$ denote an image of size $N \times N$ with $G$ gray values $(r,s)$, $(t,v) \in N \times N$, $(t,v) = (r+dx, s+dy)$ and $|.|$ is the cardinality of a set. $p(i,j)$ refers to the normalized entry of the co-occurrence matrices. That $p(i,j) = P_d(i,j)/R$, where $R$ is the total number of pixel pairs $(i,j)$. For a displacement vector $d = (dx, dy)$ and image of size $N \times M$ $R$ is given by $(N-dx)(M-dy)$. The eight nearest-neighbour resolution cells (3 by 3 matrix), which define the surrounding image pixels, were expressed in terms of their spatial orientation to the central pixel $(i, j)$ called a reference cell [3]. The eight neighbours represent all the image pixels at a distance of 1. For example, resolution cells $(i+1, j)$ and $(i-1, j)$ are the nearest neighbours to the central cell $(i, j)$ in the horizontal direction $(\theta = 0^0)$ and at a distance $(d = 1)$. This concept is extended to the three additional directions $(\theta = 45, 90, 135^0)$ as well as when a distance equals 2, 3 and so on. Of the 14 features, we found that the following 5 have the most powerful discrimination ability as texture features of the composite images: angular second moment (*ASM*), contrast (*CON*), correlation (*COR*), variance (VAR) and entropy (*ENT*). Using the formulas of the textural features, the angular second moment, contrast, correlation, variance and entropy are presented in Table 1. « Angular second moment » is a measure of homogeneity of an image. The higher value of this feature indicates that the intensity varies less in an image. « Contrast » measures local variation in an image. A high contrast value indicates a high degree of local variation. « Correlation » is a measure of linear dependency of intensity values in an image. For an image with large areas of similar intensities, correlation is much higher than for an image with noisier, uncorrelated intensities. « Variance » indicates the variation of image intensity values. For an image with identical intensity for all images, the variance would be zero and «entropy» is an indication of the complexity within an image. A complex image produces a high entropy value.

Table 1. A set of statistical features.

- The angular second moment (*ASM*)
$$ASM = \sum_i \sum_j \{p(i, j)\}^2 \qquad (2)$$

- The contrast (*CON*)
$$CON = \sum_i \sum_j (i, j)^2 \, p(i, j) \qquad (3)$$

- The correlation (*COR*)
$$COR = \frac{\sum_{i,j} ijp(i, j) - m_x . m_y}{S_x . S_y} \qquad (4)$$

- The entropy (*ENT*)
$$ENT = -\sum_i \sum_j p(i, j) \log p(i, j) \qquad (5)$$

- The variance (VAR)
$$VAR = \sum_i \sum_j (i - p(i, j))^2 \, p(i, j) \qquad (6)$$

## 2.3. Mamdani Fuzzy Inference System (FIS)

Fuzzy Inference Systems (FIS) are popular computing frameworks based on the concepts of fuzzy set theory, which have been applied with success in many fields [8]. Their success is mainly due to their closeness to human perception and reasoning, as well as their intuitive handling and simplicity, which are important factors for acceptance and usability of the systems [30]. In figure1 the general schema of a FIS is portrayed. In particular, three main modules are of particular interest: a fuzzifier, a rule base and a defuzzifier. While the fuzzifier and the defuzzifier have the role of converting external information in fuzzy quantities and vice versa, the core of a FIS is its knowledge base, which is expressed in terms of fuzzy rules and allows for approximate reasoning [28]. Typically, a FIS can be classified according to three types of models that are distinguished in the formalization of the fuzzy rules [29]. In this work, we focus on the Mamdani type, which is characterized by the following fuzzy rule schema:

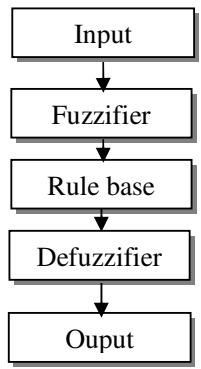

Figure 1. Scheme of a Fuzzy Inference System

The main feature of such type of FIS is that both the antecedents and the consequents of the rules are expressed as linguistic constraints [17]. As a consequence, a Mamdani FIS can provide a highly intuitive knowledge base that is easy to understand and maintain, though its rule formalization requires a time consuming defuzzification procedure. As aforementioned, the proposed adaptive watermarking scheme computes a watermark weighting function using a HVS and a FIS. In order to efficiently extract the masking information, while taking into account the local characteristics of the image. In what follows we present the insertion algorithm whose steps are detailed. In this work, the FIS is utilized to compute the optimum watermark weighting function that would enable the embedding of the maximum-energy and imperceptible watermark. This FIS is therefore ideal to model the watermark weighting function, as it can incorporate the fuzzy and nonlinear aspect of human vision. In this work, the input membership function was divided into three linguistic values, where each input denoted as low average and high respectively. The determination of the membership function is done by using the help of ANFIS Toolbox in MATLAB. This technique enabled excellent model development for non-linear process in which the rules were automatically generated under ANFIS environment.

## 3. WATERMARK EMBEDDING AND DETECTION

Here, we describe a frequency domain based watermarking system with blind detection. Basically, it comprises a watermark embedding and a detection process. Figure 2 illustrates the watermark embedding process. A host image is first transformed using one of the available transformation tools; our work in the DCT is used. Coefficients are then selected for watermark insertion. After embedding process, an inverse transformation is applied. To obtain a watermarked image. The blind detection process is shown in Figure 3. It is performed by means of a correlation function. A possibly corrupted image is transformed using the same tool as at the embedding process. Among the proposed activities in the domain of medical image watermarking, the algorithm [31], which is encoded on a pair of frequency values {0, 1}. Use of frequency domain DCT can fulfill not only the invisibility through the study of optimizing the insertion gain used, but also security by providing a blind algorithm or use the original image No is not essential and the extraction of the mark is through a secret key [5], [12].

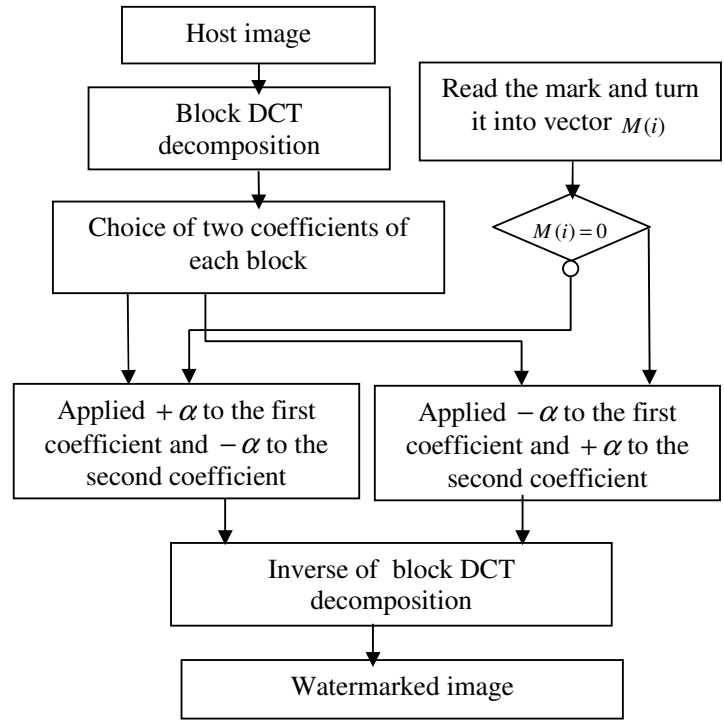

Figure 2. The watermark embedding process

In fact, the human eye is more sensitive to noise in lower frequency components than in higher frequency ones. However, the energy of most natural images is concentrated in the lower frequency range, and watermark data in the higher frequency components might be discarded after quantization operation of lossy compression. In order to invisibly embed the watermark that can survive lossy data compressions, a reasonable trade-off is to embed the watermark into the middle-frequency range of the image. In our approach we have two key insertion to secure the place where the watermark was introduced. The first key tells us the positions of the two coefficients selected with the same value of quantization [18], [20]. While the second key position Relates blocks which bear the marks among all the components total blocks transformed image. Phase extraction is as follows: Compare the values of DCT coefficients to determine if the respective bit of the message was a "0" or a "1".

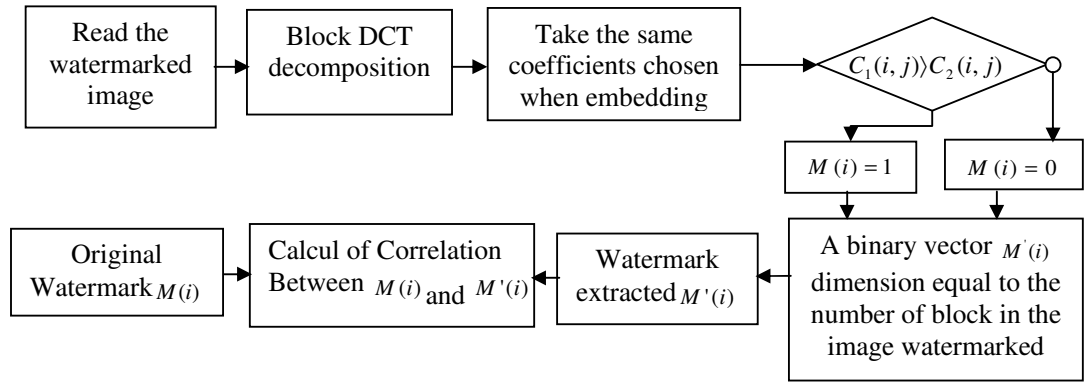

Figure 3. The watermark detection process

## 4. VALIDATION OF THE NEW APPROACH

We evaluated the performance of the proposed method by conducting several simulations that based on the algorithm described in Section3, using standard medical images. Simulation conditions, results and interpretation are presented in this section.

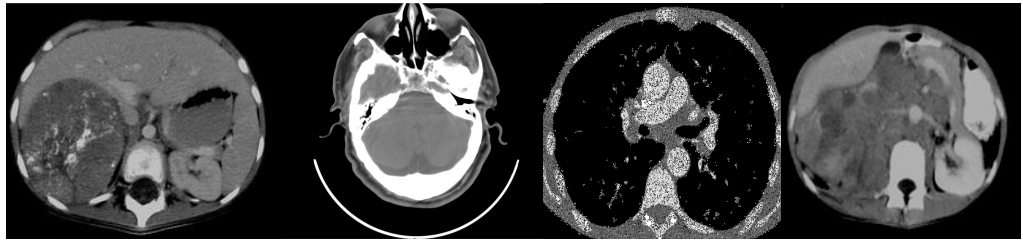

Figure 4. Originals medicals images used in experiment.

### 4.1. Robustness Measure

Nowadays, the most popular distortion measures in the image field are the Signal-to-Noise Ratio ($SNR$), and the Peak Signal-to-Noise Ratio ($PSNR$). They are usually measured in decibels, i.e., "dB":

$$MSE = \frac{1}{MN} \sum_{x=0}^{M-1} \sum_{y=0}^{N-1} (f(x,y) - f'(x,y))^2 \qquad (7)$$

Where *M, N* is the size of the image and contains *M* x *N* pixels, *f(x, y)* is the host image and *f'(x, y)* is the watermarked image. This measure gives an indication of how much degradation values near to zero indicate less degradation.

$$PSNR = 10\log_{10}(\frac{X_{max}^2}{MSE}) = 10\log_{10}(\frac{255^2}{MSE}) \qquad (8)$$

$x_{max}$ is the luminance max.

For the aim of improving the perceptual fidelity after Watermarking process, we use a new version of the Peak Signal-to-Noise Ratio as a distortion metric to estimate image degradation. It takes into consideration the importance of the human visual system and its characteristics and therefore is more suitable for digital watermarking. In fact, if the difference is located in an heterogenous region, the human eye can't distinguish it. Whereas, when the difference appears in an homogenous region, the image seems to be different from the original one. Our new watermarking scheme using the $wPSNR$ attempts to attain an optimal trade-off between estimates of robustness, data payload and perceptual fidelity. Results of experiments, carried out on a database of 256× 256 pixel-sized medical images, will demonstrate that our watermarking scheme is robust against different quality of JPEG Compression. In our paper, images are extracted from DICOM library and we use Matlab and its library as software to experiment our approaches and to validate our results. The ($wPSNR$) is used to evaluate the image quality by calculating the ($wMSE$) between the images to compare. The equations are as follows:

$$wMSE = \frac{1}{MN}\sum_{x=0}^{M-1}\sum_{y=0}^{N-1}(2*\frac{|f(x,y)-f'(x,y)|}{|f(x,y)+f(x,y)'|})^2 \qquad (9)$$

$$wPSNR = 10\log_{10}(\frac{X_{max}^2}{wMSE}) = 10\log_{10}(\frac{255^2}{wMSE}) \qquad (10)$$

By using the proposed scheme, the watermark is almost imperceptibility to the human eyes, as shown in figure 5 and figure 6.

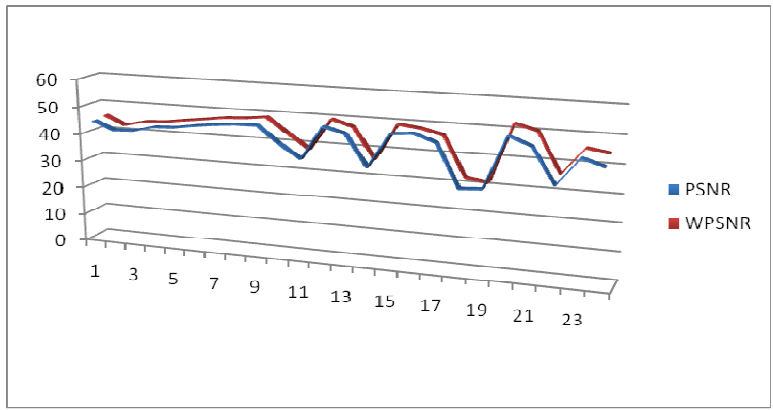

Figure 5: Mean values of $PSNR$ and $wPSNR$ per attack for different test images.

Signal & Image Processing : An International Journal(SIPIJ) Vol.1, No.2, December 2010

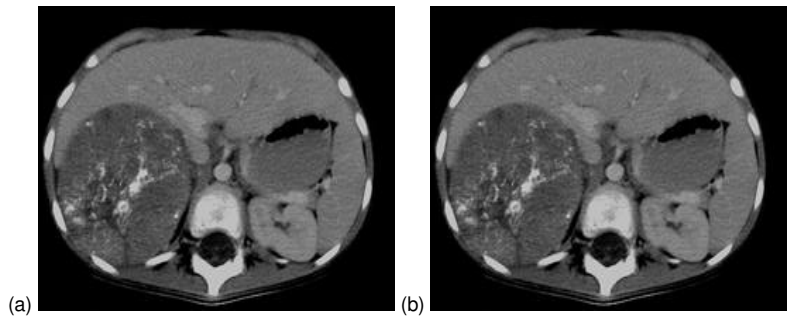

Figure 6. (a) Original medical image (b) Medical image watermarked $PSNR$ : 40.12 dB

## 4.2. Robustness towards JPEG Compression

We always need to apply JPEG compression to the medical images for archive or transmission [21]. It is therefore crucial to examine whether the proposed watermarking scheme can survive JPEG compression attacks. In the medical field, the compression rate is constantly a discussion subject because of the importance of the medical information. In order to perform this experiment, the watermarked image was compressed using different quality factors 90%, 70%, 50%, 30% and 10%. In this section we also present experimental results carried out on a database of 800 watermarks. The test technique is made by correlation between the extracted watermark and the dictionary. Figure 7 shows that the watermarks are correctly detected from our simulation results from Watermarked images after attacks.

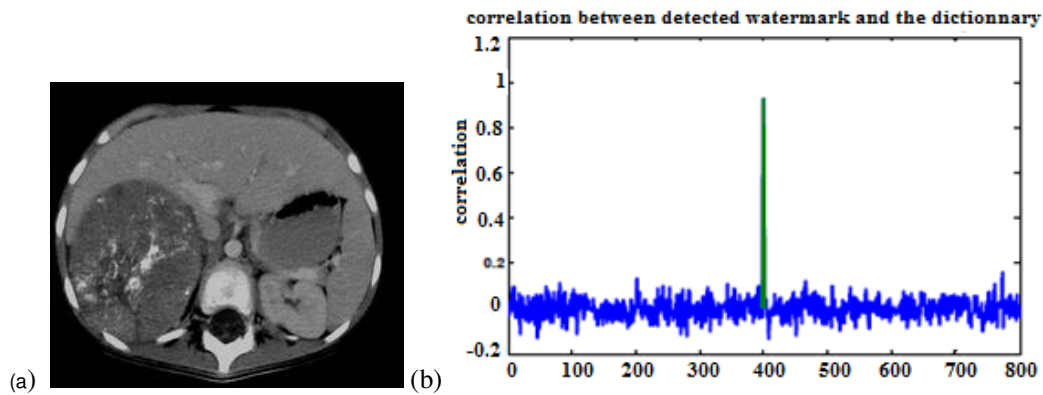

Figure 7. (a) JPEG compression (with quality factor 35%) of the watermarked image (b) Watermark detector response of attacked by JPEG compression (with quality factor 35).

We present in the following, in figure 8, the means of $PSNR$ and $wPSNR$ for the test images related to the attacks of different compression quality factors. The values of $PSNR$ an $wPSNR$ always evolve paralleling, The $wPSNR$ values are superior to those of $PSNR$ .

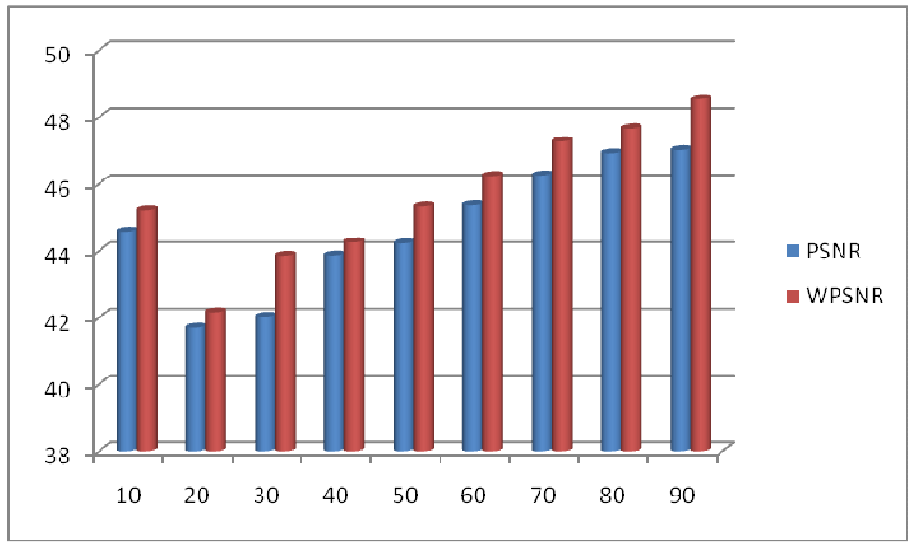

Figure 8. Means values of *PSNR* and *wPSNR* for test images watermarked and attacked by different ratio of JPEG compression.

In addition, we compare our experimental results to Cox et al.'s scheme [27], as well as Lou and Yin's scheme [25]. Watermark schemes proposed by these authors have a weakness that the HVS model has not been taken into account. For the sake of the imperceptibility constraint of a watermark, Authors include low strength watermark to avoid degrading the image quality. Unfortunately, it reduces the robustness of the watermark. In this paper, using the HVS model and the FIS, the watermark can be adjusted for each different image that provides a maximum and suitable watermark subject to the imperceptibility constraint. The results are shown in figure 9. Our method clearly outperforms the other two aforementioned schemes.

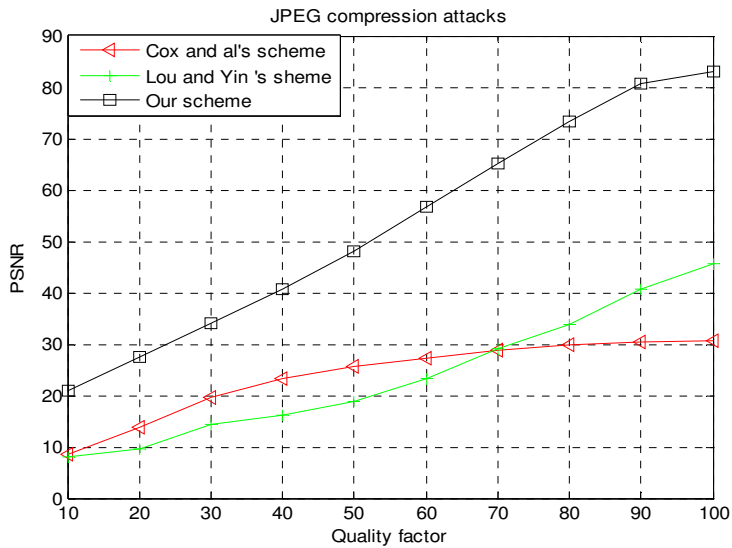

Figure 9. Comparison with different algorithm

### 4.3. Robustness towards Noise

It seems to be interesting to evaluate the robustness of the proposed method towards noise. In fact, in the medical field, the used instruments add different noise types to the medical images [14], [19]. We have tested the proposed method using different noise generations by modifying either its type or its variance. In the following scheme presents the mean values of $PSNR$ and $wPSNR$ for medical test images. The values of $PSNR$ an $wPSNR$ always evolve together. These values reach their maximum for the speckle noise. For the images attacked by a gaussian noise, the values of $PSNR$ and $wPSNR$ are minimal. Whereas, these values are always above 30 dB which allows considering that the image quality is good and these watermarked images remain useful even with the presence of these noises.

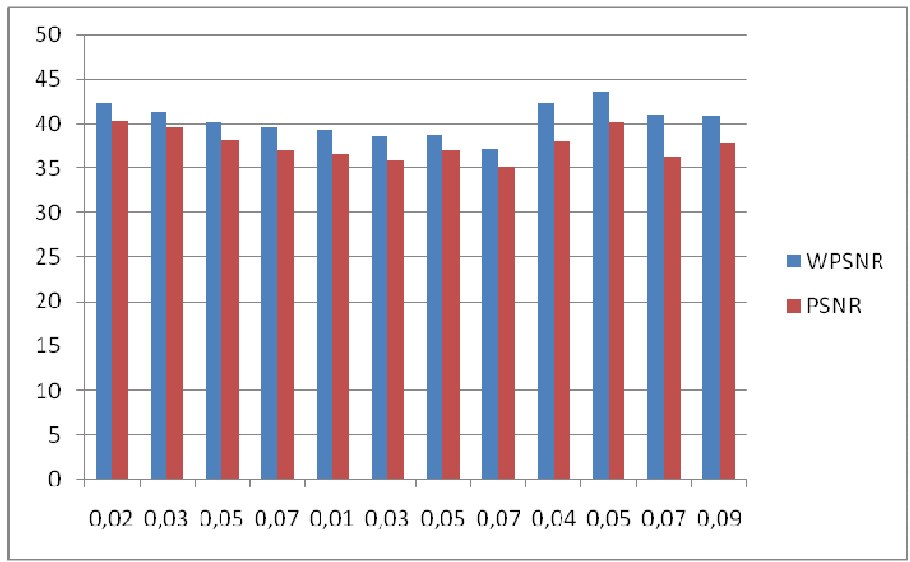

Figure 10. Means values of $PSNR$ and $wPSNR$ for several test images watermarked and attacked by different noise types.

### 4.3. Robustness towards Filtering

We have tested the robustness of our proposed method towards 3 filter types: Gaussian, unsharp and average. These are the most used filers to eliminate noise from images. We have changed the filter size until bluffing the image and we always succeeded to extract the embedded signature. Figure 11 displays good values of $PSNR$ and $wPSNR$. It follows then that fidelity in images is improved after Filtering attacks.

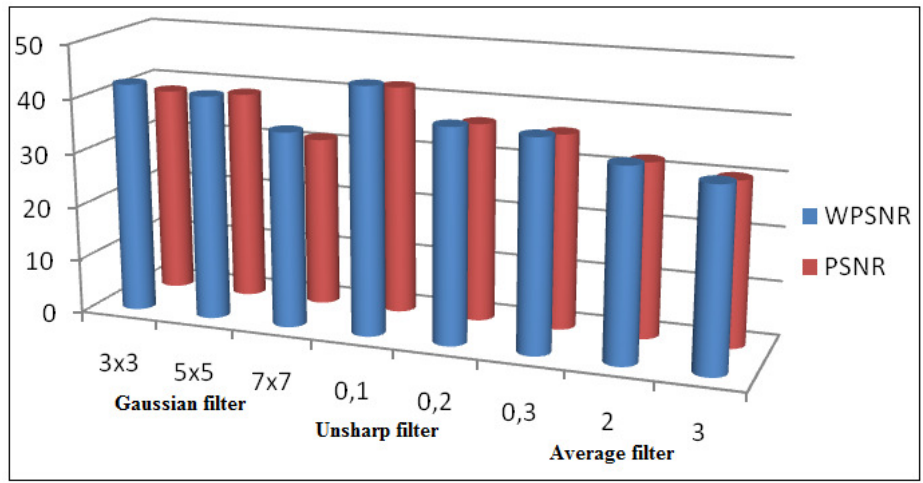

Figure 11. Mean values of *PSNR* and *wPSNR* for several tests images Watermarked and attacked by different filters

## 5. CONCLUSION

In this paper, we have presented a new approach of blind watermarking scheme used in medical images that is robust some attacks. The FIS and the HVS combined are used to adjust the watermarking strength that can be embedded without noticeably degrading the quality of the image. The watermark is embedded into the mid-band frequency range of the image after being transformed by the Discrete Cosine Transform (DCT). As a result, the watermark is more robust and imperceptible. The fuzzy inference system is able to directly provide the power level to use, instead of the current iterative process. Our current research provides a promising technique for watermarking images. As future work; we will extend the algorithm in order to obtain less degradation in the watermarked image and obtain better accuracy in the recovered watermark.

## ACKNOWLEDGEMENTS

The authors express gratitude to Dr. REZGUI Haythem and Dr. AZAIEZ Mustapha clinic from the MANAR II of Tunisia, for helping and being in assistance.

**Authors**

**S**. **Oueslati** is a researcher at the Image and Information Processing Department Higher National School of Telecommunications of Bretagne she is also in signal processing laboratory at the University of Sciences of Tunis - Tunisia (FST). Degree in electronics and she received a Masters degree in 2006 from the University of Sciences of Tunis.  She is currently a PhD student at the Faculty of Sciences of Tunis of where she is a contractual assistant. His research interests include information hiding and image processing, digital watermarking, database security.

**A.Cherif** obtained his engineering diploma in 1988 from the Engineering Faculty of Tunis and his Ph.D. in electrical engineering and electronics in 1997. Actually he is a professor at the Science Faculty of Tunis, responsible for the Signal Processing Laboratory. He participated in several research and cooperation projects, and is the author of more than 60 international communications and publications.

**B. Solaiman**, Ing. Ph.D., Professor, is Telecom Engineer, holds a Ph.D. and HDR in Information Processing, University of Rennes I, is currently Professor and Head of Image and Information Processing from the Higher National School of Telecommunication of Bretagne in Brest, France. His research interests include, among others, on different approaches to treatment and Information Fusion and have been the subject of numerous publications.